\documentclass{elsart}

\usepackage{graphicx}

\usepackage{amssymb}

\newcommand{\kb}{k_{\rm B}}
\newcommand{\cL}{\mathcal{L}}
\newcommand{\ccdot}{\!\cdot\!}                       

\newcommand{\bD}{\mathbf{D}}
\newcommand{\bh}{\mathbf{h}}
\newcommand{\bH}{\mathbf{H}}
\newcommand{\bM}{\mathbf{M}}
\newcommand{\bn}{\mathbf{n}}

\newcommand{\bu}{\mathbf{u}}
\newcommand{\bv}{\mathbf{v}}

\newcommand{\bone}{\mathbf{1}}

\newcommand{\bxi}{\mbox{\boldmath${\xi}$}}
\newcommand{\bOmega}{\mbox{\boldmath${\Omega}$}}
\newcommand{\bDelta}{\mathbf{\Delta}}
\newcommand{\bUt}{\mbox{\boldmath$U$}_{\!t}}
\newcommand{\bWt}{\mbox{\boldmath$W$}_{\!t}}
\newcommand{\UP}{\mbox{\boldmath$P$}_{\!t}}          

\newcommand{\setm}{\mathbf{m}}
\newcommand{\setM}{\mathbf{M}}

\begin{document}

\begin{frontmatter}


\title{Combined micro--macro integration scheme from an invariance principle:
application to ferrofluid dynamics}
\author[label1,label2]{Patrick Ilg}
\address[label1]{Inst. Theoret. Physik, TU Berlin, Sekr. PN 7-1, Hardenbergstr. 36,
D-10623 Berlin, Germany}
\ead{ilg@itp.physik.tu-berlin.de}
\thanks[label2]{corresponding author; phone: +49-30 314 25225; fax: +49-30 314 21130}
\author[label3,label4]{Iliya V. Karlin}
\address[label3]{Department of Materials, Institute of Polymers, ML J19,
ETH Z{\"u}rich, CH-8092 Z{\"u}rich, Switzerland}
\address[label4]{Institute of Computational Modelling SB RAS, 660036 Krasnoyarsk, Russia}
\ead{ikarlin@mat.ethz.ch}

\title{}


\author{}

\address{}

\begin{abstract}
A method for the combination of microscopic and macroscopic
simulations is developed which is based on the invariance of the
macroscopic relative to the microscopic dynamics. The method
recognizes the onset and breakdown of the macroscopic description
during the integration. We apply this method to the case of
ferrofluid dynamics, where it switches between direct Brownian
dynamics simulations and integration of the constitutive equation.
\end{abstract}

\end{frontmatter}

\section{Introduction}
\label{intro} The understanding of the macroscopic dynamics from
the underlying microscopic time evolution is a central issue of
non--equilibrium statistical mechanics. The massive use of
computer simulations over the last years has led to new approaches
to this very old problem. Among others, we mention Legendre
integrators \cite{Ilg_RedFENE,Ilg_RedLCP,GoGoKa03}, the CONNFESSIT
method \cite{connfess}, adaptive mesh refinement and multiscale
modeling \cite{Jendrejack03,Kevrekidis03}. The last two methods do
not require knowledge of the macroscopic equations. On the other
hand, there has been much effort to derive approximate macroscopic
equations from the microscopic dynamics, which yield reliable
results under various circumstances (see e.g.~Ref.~\cite{BiWi95}
for an overview of constitutive equations for polymer liquids).
Here, we follow the approach proposed in Ref.~\cite{GKILOe01} to
combine microscopic and macroscopic simulations in a combined
integration scheme which recognizes the onset and breakdown of the
chosen macroscopic description during the simulation. Note, that
the breakdown of a chosen macroscopic description does not imply a
similar breakdown of other, improved macroscopic descriptions.
Instead of improving the macroscopic equations, which is the aim
of many works on closure approximations (see
e.g.~\cite{Ilg_RedFENE,Ilg_RedLCP} and references therein), we here keep the
chosen macroscopic description and use it as long and as
frequently in the simulation as possible. While this integration
scheme was used in Ref.~\cite{GKILOe01} to detect the onset of the
macroscopic description, we here present the full scheme that
switches back and forth between microscopic and macroscopic
simulations. We apply this scheme to well--known models of
ferrofluid dynamics where it decides between direct Brownian
dynamics simulations and integration of the constitutive equation.

\section{Invariance principle and combined integration scheme}
\label{HybridScheme} In order to keep the paper self--contained,
we briefly summarize the main ideas of the combined integration
scheme based on the invariance principle proposed in
Refs.~\cite{GKILOe01,Iswitch02}. We assume a given microscopic
description of the system, where the microscopic variables are
denoted by $f$. The microscopic dynamics is specified by the
vector field $J$,
\begin{equation} \label{kinetic}
        \frac{\partial}{\partial t} f = J(f).
\end{equation}
In addition, we assume that the set of macroscopic variables
$\setM$ has been chosen. Typically, $f(x)$ is the distribution function over
the set of microscopic coordinates $x$ and $\setM$ contains low--order
moments of $f$. In this case, the macroscopic variables are linear
functionals of the microscopic distribution function,
$\setM=\int\!dx\,\setm(x)f(x)$.
Although the method can be applied to more general situations,
we here limit ourselves to this case for the sake of clarity.

The reduced or macroscopic description assumes not only closed--form
macroscopic equations $\dot{\setM}(\setM)$, but also a family of
canonical distribution functions $f_{\setM}(x)$
\cite{Ilg_RedFENE,Ilg_RedLCP}.
The canonical distribution functions satisfy the consistency
relation
$\setM=\int\!dx\,\setm(x)f_{\setM}(x)$.
Then, the macroscopic dynamics is given by
\begin{equation} \label{dtmacro}
        \dot{\setM}(\setM) = \int\!dx\,\setm(x)J(f_{\setM}(x))
\end{equation}
Different routes to the construction of
$f_{\setM}$ have been proposed.
In many applications, the dynamic system Eq.~(\ref{kinetic})
is equipped with a Lyapunov function $S$
(the entropy, free energy, etc.),
and the canonical distribution functions $f_{\setM}$
are conditional maximizers of $S$ subject to fixed $\setM$
\cite{GKILOe01}.

In order to estimate the accuracy of the macroscopic description
we define the defect of invariance $\Delta_{\setM}(x)$ as the difference of
the microscopic and macroscopic time derivative,
\begin{equation} \label{defect}
        \Delta_{\setM}(x) =
        J(f_{\setM}) -
        \frac{\partial f_{\setM}}{\partial \setM}\cdot\dot{\setM}(\setM).
\end{equation}
By construction, $\int\!dx\,\setm(x)\Delta_{\setM}(x)=0$.
If the defect of invariance $\Delta_{\setM}(x)$ vanishes for all
admissible values of
$\setM$, then the reduced description is called invariant and the
family $f_{\setM}$ represents the invariant manifold in the space of
the microscopic variables. The invariant manifold is relevant if it
is stable. Exact invariant manifolds are known only in very
few cases.
Corrections to the manifold $f_{\setM}$ through minimizing $\Delta_{\setM}$
is part of the so--called method of invariant manifolds
\cite{GK92,GKZinovyev03}.

Here, we exploit the invariance principle in a different way.
Let $f(x;t|t_0)$ denote the microscopic variables at time $t$ for
given initial conditions at time $t_0<t$. The values of the macroscopic
variables at time $t$ are given by
$\setM(t|t_0)=\int\!dx\,\setm(x)f(x;t|t_0)$.
On the other hand, the solution of the macroscopic equations
(\ref{dtmacro})  with
corresponding initial conditions gives $\setM^\ast(t|t_0)$.
We denote with $\|\Delta\|$ the value of the defect of invariance
(\ref{defect}) with respect to some norm $\|\bullet\|$
and $\epsilon>0$ a fixed threshold value.
If at time $t$ the defect of invariance satisfies
\begin{equation} \label{onset}
        \|\Delta_{\setM(t|t_0)}\| < \epsilon,
\end{equation}
it is said that the macroscopic description {\em sets on},
since the reduced description is sufficiently accurate.
However, if
\begin{equation} \label{breakdown}
        \|\Delta_{\setM^\ast(t|t_0)}\| > \epsilon,
\end{equation}
the macroscopic description {\em breaks down} since
the accuracy of the macroscopic dynamics is insufficient.
Therefore, the evaluation of the defect of invariance (\ref{defect})
on the current solution either to the macroscopic or to the
microscopic dynamics and checking Eqs.~(\ref{onset}) and (\ref{breakdown})
we can decide whether integration of the macroscopic dynamics is
sufficiently accurate or not.

This information is used in the combined integration scheme to
switch between microscopic and macroscopic simulations. The scheme
is sketched in Fig.~\ref{fig_hybridscheme}. Suppose at time $t_0$
the microscopic dynamics is integrated for given initial
condition. The integration is continued until at time $t_1$ the
inequality (\ref{onset}) is satisfied. At this point, the
macroscopic dynamics is started with the actual values of the
macroscopic variables, $\setM^\ast(t_1)=\setM(t_1|t_0)$. The
macroscopic dynamics is integrated until the macroscopic
description breaks down at a later time $t_2$, which is signaled
by $\|\Delta_{\setM^\ast(t_2|t_1)}\|>\epsilon$. At this time it is
necessary to switch back from the macroscopic to the microscopic
simulations in order to achieve the required accuracy. The initial
condition for the microscopic simulation at time $t_2$ is obtained
from the macroscopic description, $f(x,t_2)=f_{\setM(t_2)}(x)$.
Then, the microscopic dynamics is integrated until the macroscopic
description sets on etc. In the sequel, we demonstrate this scheme
for the case of ferrofluid dynamics.

\section{Kinetic models of ferrofluid dynamics}
\label{ferrofluids}
Ferrofluids are stable suspensions of nano--sized ferromagnetic
colloidal particles in a suitable carrier liquid
\cite{BLUMS97}.
These fluids attract considerable interest due to their peculiar
behavior, such as the magnetoviscous effect, the dependence of
the viscosity coefficients on the magnetic field
\cite{KrogerIlgHess_JPHYS03}.

We here consider the kinetic model of ferrofluid dynamics proposed in
Refs.~\cite{Martensyuk73,Cebers84b}.
In this model, the ferromagnetic particles are assumed to be
identical, magnetically hard ferromagnetic monodomain particles.
It is further assumed that the particles are of an
ellipsoidal shape with axes ratio $r$ and that
the magnetic moment $\mu$ is oriented parallel to the symmetry axes
of the particle.
Let $f(\bu)$ denote the orientational distribution
function to find a ferromagnetic particle with the orientation $\bu$,
where $\bu$ is a vector on the three-dimensional unit sphere.
In the general notation of Sec.~\ref{HybridScheme}, the microscopic
coordinates $x$ are the orientations $\bu$ and $f(x)$ is the
orientational distribution function $f(\bu)$.
The normalized macroscopic magnetization $\bM$ is given by
$\bM = \int\!d^2u\,\bu f(\bu)$, where
$\int\!d^2u$ denotes integration over the three--dimensional unit
sphere. The normalization is performed with the saturation
magnetization $M_{\rm sat}=n\mu$, where $n$ denotes the number density.
In the presence of a local magnetic field $\bH$ and a velocity field
$\bv$, the dynamics is given by \cite{BLUMS97,Hess76}
\begin{equation} \label{FFkinetic}
        \frac{\partial}{\partial t} f =
                -\cL\cdot\{(\bOmega\times\bu+B\bu\times\bD\cdot\bu
                - D_r \bh\times\bu )f\}
                +D_r \cL\cdot\cL f
\end{equation}
In Eq.~(\ref{FFkinetic}) we have introduced the rotational operator
$\cL=\bu\times\partial/\partial\bu$,
the vorticity
$\bOmega=\nabla\times\bv/2$, the symmetric velocity gradient
$2\bD=\nabla\bv + (\nabla\bv)^{\rm T}$,
and the so--called shape factor $B=(r^2-1)/(r^2+1)$.
The rotational diffusion coefficient $D_r$ defines the rotational
relaxation time $\tau=(2D_r)^{-1}$.
The dimensionless magnetic field is defined by $\bh=\mu\bH/\kb T$,
where $\kb$ and $T$ denote Boltzmann's constant and temperature,
respectively.
The equilibrium distribution
\begin{equation} \label{f_eq}
        f_{\bh}(\bu) = \frac{h}{4\pi\sinh(h)}\exp{[\bh\cdot\bu]}
\end{equation}
is the stationary solution to the
kinetic equation (\ref{FFkinetic}) in the absence of flow.
From Eq.~(\ref{f_eq}), the equilibrium magnetization
$\bM^{\rm eq}$ is found to be given by
$\bM^{\rm eq}/M_{\rm sat} = L_1(h)\bh/h$, where $h=\sqrt{\bh\cdot\bh}$
is the Langevin parameter and
$L_1(x)=\coth(x)-x^{-1}$ denotes the Langevin function.

Except for special cases, exact solutions to the kinetic equation
(\ref{FFkinetic}) are unknown and closed form equations for the
magnetization cannot be derived exactly from
Eq.~(\ref{FFkinetic}). In order to solve the closure problem, the
authors of Ref.~\cite{MRS74} have suggested to use the family of
equilibrium distributions (\ref{f_eq}), where the magnetic field
$\bh$ is replaced by an effective field $\bxi$. Thus, the
non--equilibrium magnetization is given by $\bM/M_{\rm sat} =
L_1(\xi)\bn$, where we have introduced the norm $\xi$ of the
effective field $\bxi$ and $\bn=\bxi/\xi$. This so--called
Effective Field Approximation (EFA) is a particular instance of
the quasi-equilibrium or maximum entropy approximation. It is
derived from extremizing the entropy functional
$S[f]=-\int\!d^2u\,f(\bu)\ln[f(\bu)/f_{\bh}(\bu)]$ subject to the
constraints of fixed normalization and fixed values of
magnetization $\bM$. Therefore, the set of macroscopic variables
contains only the magnetization $\bM$ in the present case. For
more details on the use of quasi-equilibrium approximation in the
context of complex fluids see e.g.
\cite{Ilg_RedFENE,Ilg_RedLCP,GoGoKa03}. The macroscopic equation
(\ref{dtmacro}) becomes the magnetization equation
\begin{eqnarray} \label{dt_u}
        \dot{\bM} & = & \bOmega\times\bM +
        B\left[\left( 1- \frac{2L_2(\xi)}{\xi L_1(\xi)} \right)\bD\cdot\bM
        -\frac{L_3(\xi)}{L_1(\xi)}(\bD\colon\bn\bn)\bM\right] \nonumber\\
        && - 2D_r\bM +
        D_r\left[ \left(1-\frac{L_1(\xi)}{\xi}\right)\bh -
        \frac{L_2(\xi)}{L_1(\xi)}(\bh\cdot\bn)\bM\right].
\end{eqnarray}
Functions $L_i(x)$ are defined recursively by
$L_{i+1}(x)=L_{i-1}(x) - (2i+1)L_i(x)/x$, with $L_0(x)=1$ and
$L_1(x)$ the Langevin function.
The accuracy of the EFA has been discussed,
e.g., in Refs.~\cite{MRS74,IKH01,IK02,IKHZ03}.

The defect (\ref{defect}) of this approximation can be calculated
explicitly.
Note, that $\int\!d^2u\,\Delta_{\setM}$ as well as
$\int\!d^2u\,\bu\Delta_{\setM}$
vanish identically by construction.
Therefore, we estimate the accuracy of the EFA through
information about the dynamics of the next higher order moment that is
not included in the macroscopic description and consider
the matrix $\bDelta_{\setM}=\int\!d^2u\,\bu\bu\Delta_{\setM}$.
As a suitable norm we use the matrix norm
$\|\Delta\|=\sqrt{\sum_{\alpha\beta}(\Delta_{\alpha\beta})^2}$.
The matrix $\bDelta_{\setM}$ can be represented by
\begin{equation} \label{uudefect}
        \bDelta_{\setM} = d_1 \bone + d_2 \bn\bn + d_3 (\bh\bn+\bn\bh) +
        d_4 \bD + d_5 (\bD\cdot\bn\bn + \bn\bn\cdot\bD),
\end{equation}
where the coefficients $d_i$ are defined in the Appendix
\ref{appendCoeffs}.

\section{Numerical implementation}
\label{numerics}
The microscopic dynamics (\ref{FFkinetic}) is integrated by
Brownian dynamics simulations of the corresponding It\^o
stochastic differential equation
\begin{eqnarray} \label{sde}
        d\bUt \!&=&\!
        \UP \!\cdot\!\left[
        \left(\bOmega\!\times\!\bUt +
          B \bD\ccdot\bUt
        + \bh \right) dt
        + d\bWt \right]
        \!- \bUt \frac{dt}{\tau}.
\end{eqnarray}
The projector perpendicular to $\bUt$ is denoted by
$\UP\equiv (\bone-\bUt\bUt)$
and $\bWt$ is a three--dimensional Wiener process
\cite{HCO96}.
Using It\^o's formula,
it is verified that Eq.~(\ref{sde}) conserves the normalization of
$\bUt$.
Eq.~(\ref{sde}) is integrated numerically by
a weak first-order scheme that guarantees the normalization
of the random unit vector $\bUt$ \cite{HCO96}.
An ensemble of $10^5$ random vectors $\bUt$ is used in the simulation
in order to ensure accurate ensemble averages.
A constant time step $\Delta t/\tau=10^{-3}$ is used throughout.

The macroscopic equation (\ref{dt_u}) is integrated
directly in terms of the macroscopic variables $\bM$.
The effective field $\xi$ is calculated by
$\xi=L_1^{-1}(M)$, where $M$ denotes the norm of $\bM$ and
$L_1^{-1}(x)$ denotes the inverse Langevin function.
The latter is evaluated numerically by the Newton--Raphson
Method.
In order to treat microscopic and macroscopic dynamics approximately
on the same footing, we used a first-order explicit Euler scheme
with the same time step $\Delta t/\tau=10^{-3}$ to integrate
Eq.~(\ref{dt_u}) numerically.
More accurate schemes for the macroscopic equation,
such as Runge--Kutta methods, could be employed as well.

At every time step of the microscopic or macroscopic integration,
the norm of the defect of invariance $\|\Delta\|$ is evaluated
from Eq.~(\ref{uudefect}). If in the course of the microscopic
integration the inequality $\|\Delta\|<\epsilon$ is satisfied at
time $t$, then the microscopic simulation is stopped and the
macroscopic integration of Eq.~(\ref{dt_u}) is started with the
initial condition $\bM(t)$ calculated from the microscopic
ensemble average. On the other hand, if at time $t$ during the
integration of the macroscopic equation $\|\Delta\|>\epsilon$ is
fulfilled, the microscopic simulation is started with initial
condition $f(\bu;t)=f_{\setM(t)}(\bu)$. Here, we use the rejection
method \cite{Recipes} in order to generate an ensemble of random
unit vectors $\bUt$, which are distributed according to
$f_{\setM(t)}(\bu)$.

\section{Results}
\label{result} The combined integration scheme described above has
been implemented and run for a number of different values of
magnetic field and velocity gradients. We limited ourselves to the
case of plane shear flow. Previous investigations showed that the
EFA provides a good approximation to the kinetic equation
(\ref{FFkinetic}) for moderate values of axes ratios $r$
\cite{IKHZ03}. We here choose a value of $r=5$ in the sequel,
where the accuracy of the EFA is not as good as for smaller
values.

First, we consider the dynamics in the absence of any velocity
gradients where a constant magnetic field $h$ is applied during
the time interval $t_{\rm i}\leq t \leq t_{\rm f}$. Comparison of
the BD simulation results and the EFA for $h=2$ and $h=5$ with
$t_{\rm i}=\tau$, $t_{\rm f}=4\tau$ is shown in
Fig.~\ref{fig_stepH_BDvsEFA}. From Fig.~\ref{fig_stepH_BDvsEFA} we
observe that the EFA provides a very good approximation for this
case. Deviations of the EFA from the results of the BD simulation
are shown in Fig.~\ref{fig_diffanddefect_stepH} together with the
values of the norm of the defect of invariance. As can be seen
from Fig.~\ref{fig_diffanddefect_stepH}, the defect is very
sensitive to deviations of the EFA from the BD results. Upon
closer inspection one recognizes from
Fig.~\ref{fig_diffanddefect_stepH} that the maxima of the defect
preceed the extrema of the deviations, reflecting the fact that
the defect of invariance is sensitive to deviations in the time
derivative. Thus, the defect signals the inaccuracy of the time
derivative of the EFA and allows for a switch to the BD simulation
before the values of the magnetization become inaccurate. The
result of the combined integration scheme with $\epsilon=0.2$ is
shown in Fig.~\ref{fig_stepH_hybridvsBDvsEFA}. Within the boxed
regions, the norm of the defect of invariance
$\|\Delta\|>\epsilon$ and the BD simulation is performed while
otherwise the EFA is integrated. In the inset of
Fig.~\ref{fig_stepH_hybridvsBDvsEFA}, the result of the EFA,
Fig.~\ref{fig_stepH_BDvsEFA}, is shown for comparison. Clearly,
the combined integration improves the comparison of the EFA to the
BD simulation.

Next, we consider the magnetization dynamics in the presence of
flow. Starting with an isotropic initial distribution in the
absence of magnetic fields and velocity gradients, a constant
magnetic field $h$ is applied during the interval $t_{\rm i}\leq t
\leq t_{\rm f}$, while the magnetic field is absent outside this
interval. In addition, a plane Couette flow with velocity field
$\bv=(2\dot{\gamma}y,0,0)$ with a constant shear rate
$\dot{\gamma}$ is applied during the interval $t_1\leq t\leq t_2$
while no shear is applied for $t<t_1$ and $t>t_2$.
Fig.~\ref{fig_stepHstepshear_BDvsEFA} shows the magnetization
dynamics for $h=5$, $\tau\dot{\gamma}=2$, $t_{\rm i}=\tau$,
$t_{\rm f}=6\tau$, $t_1=3\tau$, and $t_2=8\tau$. One observes from
Fig.~\ref{fig_stepHstepshear_BDvsEFA} that the EFA predicts
qualitativly correct behavior but fails to give accurate results
as long as the shear flow is applied.
Fig.~\ref{fig_diffanddefect_stepHstepshear} shows the deviation of
the EFA from the BD simulation results together with the norm of
the defect of invariance. Except for initial transient dynamics
when the magnetic field is suddenly applied (see
Fig.~\ref{fig_diffanddefect_stepH}) the EFA is accurate in the
absence of shear flow and $\|\Delta\|$ is small. During
application of the shear flow, however, the predictions of the EFA
are less accurate as is monitored by $\|\Delta\|$. In
Fig.~\ref{fig_stepHstepshear_hybridvsBD} we show the result of the
combined integration scheme for the present situation where
$\epsilon=0.2$. The agreement between the combined scheme with the
BD simulation is very good. Due to the low accuracy of the EFA,
the macroscopic equation (\ref{dt_u}) is integrated only in the
boxed regions while otherwise BD simulations are performed.

Finally, we consider the magnetization dynamics in a constant
magnetic field $h$ and a plane Couette flow
$\bv=(2\dot{\gamma}y,0,0)$ with oscillatory shear rate
$\dot{\gamma}=\omega\gamma_0\cos(\omega t)$. The magnetic field is
oriented in the flow direction. Equilibrium initial conditions are
chosen. Due to the flow, a nonequilibrium magnetization component
$M_y$ arises which oscillates with frequency $\omega$, while the
magnetization component in the magnetic field direction $M_x$
oscillates with frequency $2\omega$. From
Fig.~\ref{fig_oscillshear_BDvsEFA} we observe that the predictions
of the EFA are less reliable compared to the situation without
flow. The deviation of both magnetization components are shown in
Fig.~\ref{fig_diffanddefect_oscillshear} together with the defect
of invariance. The deviation of the magnetization components of
the macroscopic dynamics (EFA) from the microscopic dynamics (BD)
are seen to oscillate as well. While the deviation of $M_y$
oscillates around zero, the macroscopic dynamics always
overpredicts the values of $M_x$ in the present case. The
oscillations in the deviation of $M_x$ and $M_y$ are reflected in
the oscillations of $\|\Delta\|$. As before, the maxima of
$\|\Delta\|$ occur before the maximum deviations of the
magnetization are observed. Note that $\|\Delta\|$ always exceeds
a certain value $\epsilon_0$, which signals the inaccuracy of the
EFA in the present case. Fig.~\ref{fig_oscilshear_hybridvsBDvsEFA}
shows the result of the combined integration scheme with
$\epsilon=0.2$ for the same conditions as in
Fig.~\ref{fig_oscillshear_BDvsEFA}. In the boxed regions
$\|\Delta\|<\epsilon$ and the macroscopic (EFA) dynamics is
integrated while otherwise BD simulations are performed. The
agreement between combined integration and BD simulation is very
good. However, due to the limited accuracy of the EFA for this
case only a limited fraction of the total integration time is
preformed with the EFA.
Fig.~\ref{fig_oscilshear_Tsteop_hybridvsBDvsEFA} shows the same
situation as in Fig.~\ref{fig_oscillshear_BDvsEFA}, but the shear
flow was stopped at time $t=6.5\tau$ while the magnetic field
remained unchanged. Without the terms arising from the velocity
gradient, $\|\Delta\|$ drops below the threshold value $\epsilon$
and, within the combined scheme, the relaxational dynamics for
$t>6.5\tau$ is integrated by the EFA. Good agreement with the
result of full BD simulation is found.

\section{Conclusions}
\label{end} The present approach of combined microscopic and
macroscopic simulation exploits the invariance of the microscopic
relative to the macroscopic dynamics. It is applicable whenever a
macroscopic description to an underlying microscopic dynamics is
given. While previous studies considered the case of dilute
polymer solutions \cite{GKILOe01,Iswitch02}, the combined
integration scheme is illustrated here for the case of ferrofluid
dynamics.

The microscopic dynamics is integrated only if the defect of
invariance exceeds a certain threshold value $\epsilon$. The full
microscopic simulation is recovered for $\epsilon=0$ while
$\epsilon\to\infty$ corresponds to the macroscopic dynamics. Thus,
the combined integration improves the accuracy of the macroscopic
description, where the improvement depends on the choice of
$\epsilon$. At the same time, the combined integration saves CPU
time since the macroscopic simulation is employed whenever
possible.
The amount of CPU time that can be saved by the combined
integration scheme for a given value of $\epsilon$ depends on the
quality of the macroscopic description for the corresponding
situation. For the conditions of
Fig.~\ref{fig_oscillshear_BDvsEFA}, a sample study of the relative
error $R_x$ of the magnetization $M_x$ as a function of elapsed
CPU time is shown in Fig.~\ref{fig_error_oscilshear}. The relative
error with respect to the result of the BD simulation is defined
as $R^2_x=\frac{1}{N_t}\sum_j^{N_t}[(M_x(t_j)-M_x^{\rm
BD}(t_j))/M_x^{\rm BD}(t_j)]^2$, where $N_t$ denotes the total
number of integration time steps. For a better comparison, all
data shown in Fig.~\ref{fig_error_oscilshear} are obtained with
the same PC with a P4 processor. From
Fig.~\ref{fig_error_oscilshear} we observe that the relative error
decreases with decreasing $\epsilon$ while the time the
microscopic simulation is integrated in the combined scheme
increases and thus the required CPU time increases. Overall, we
observe that the relative error decreases almost linearly with
elapsed CPU time. Note, that $R_x=0$ does not correspond to the
exact result but to the BD simulation.

\section*{Acknowledgment}
Valuable discussions with A.~N.~Gorban, H.C.~{\"O}ttinger, and S.~Hess
are gratefully acknowledged.
This work was supported in part by DFG priority program SPP 1104
'Colloidal Magnetic Fluids' under grant no.~HE1100/6-2.

\begin{appendix}
\section{Coefficients of defect of invariance}
\label{appendCoeffs}
The coefficients $d_i$ in Eq.~(\ref{uudefect}) contain contributions
from Brownian motion, the magnetic field and the symmetric velocity
gradient.
In particular,
\begin{equation}
        d_1 = 2(L_2(\xi)+c(\xi))\left( 1- \frac{\bn\cdot\bh}{\xi}\right)
              -B\left( 2\frac{L_3(\xi)}{\xi}+\frac{3L_2(\xi)}{\xi L_1(\xi)}c(\xi)\right)(\bD\colon\bn\bn)
\end{equation}
\begin{eqnarray}
        d_2 & = & -6(L_2(\xi)+c(\xi))\left( 1- \frac{\bn\cdot\bh}{\xi}\right)
        +2\left(L_2(\xi)[\frac{1}{L_1(\xi)}-\frac{4}{\xi}]-L_3(\xi)\right)\bn\cdot\bh \nonumber\\
        & &{} + \frac{B}{\xi}\left( 14L_3(\xi)-4\frac{L_2^2(\xi)}{L_1(\xi)}
        +9\frac{L_2(\xi)}{L_1(\xi)}\right)\bD\colon\bn\bn
\end{eqnarray}
\begin{equation}
        d_3 = L_1(\xi) - L_2(\xi)\left( \frac{1}{L_1(\xi)}-\frac{1}{\xi} \right)
\end{equation}
\begin{equation}
        d_4 = \frac{2B}{\xi}\left( L_1(\xi) - \frac{2L_2(\xi)}{\xi} \right)
\end{equation}
\begin{equation}
        d_5 = \frac{2B}{\xi}\left( \frac{L_2^2(\xi)}{L_1(\xi)} - 2L_3(\xi)\right),
\end{equation}
where
\begin{equation}
        c(\xi)=\frac{L_1(\xi)}{\xi L_1'(\xi)}[L_2(\xi)-L_1(\xi)^2]
\end{equation}
and the total derivative of the Langevin function can be expressed by
\begin{equation}
       L_1'(\xi) \equiv \frac{dL_1(\xi)}{d\xi} =
       1 - \frac{2L_1(\xi)}{\xi} - L_1^2(\xi).
\end{equation}
\end{appendix}

\clearpage


\clearpage

\section*{Figure captions:}
\begin{description}
\item{\bf Fig. 1}
          Sketch of the combined integration scheme.
          The macroscopic dynamics is integrated whenever the
          norm of the defect of invariance $\|\Delta\|$ is smaller
          than some fixed threshold value $\epsilon$. Otherwise,
          the microscopic dynamics is integrated.
\item{\bf Fig. 2}
        Magnetization dynamics as a function of reduced time $t/\tau$
        in the absence of velocity gradients.
        A constant magnetic field $h$ was applied during the time
        interval $1\leq t/\tau\leq 4$, while the magnetic field was
        switched off outside this interval.
        Circles and squares are the results of the BD simulation for
        $h=5$ and $h=2$, respectively, while solid and dashed line
        are the corresponding predictions of the EFA.
        The inset shows the comparison for $h=5$ on a finer scale.
\item{\bf Fig. 3}
  Deviation of normalized magnetization $M_x/M_{\rm sat}$ calculated
        from BD simulation and the EFA for $h=5$ (circles) and
        $h=2$ (squares) for the same conditions as in
        Fig.~\ref{fig_stepH_BDvsEFA}.
        Solid and dashed lines are the defect of invariance
        as calculated from the matrix norm of Eq.~(\ref{uudefect}) for
        $h=5$ and $h=2$, respectively. For better visibility,
        the matrix norm was multiplied by a factor $0.1$.
\item{\bf Fig. 4}
        Magnetization dynamics as a function of reduced time $t/\tau$
        for the same condition as in Fig.~\ref{fig_stepH_BDvsEFA}.
        Circles and squares are the result of the BD simulation for
        $h=5$ and $h=2$, respectively, while solid and dashed lines
        are the results of the combined integration scheme with
        $\epsilon=0.2$ for $h=5$ and $h=2$, respectively.
        Within the boxed regions (indicated by the shading in the upper 
        part), $\|\Delta\|>\epsilon$ and the
        BD simulation is performed, otherwise the EFA is integrated.
        The inset shows the comparison for $h=5$ on a finer scale,
        where the dashed--dotted line is the result of the EFA.
\item{\bf Fig. 5}
        Magnetization dynamics as a function of reduced time
        $t/\tau$ in a constant magnetic field $h=5$ for
        $1\leq t/\tau\leq 6$ and steady shear flow with shear rate
        $\tau\dot{\gamma}=2$ for $3\leq t/\tau\leq 8$,
        where the magnetic field is oriented in the gradient direction.
        No magnetic field and no shear flow is applied outside the
        mentioned time intervals.
        Circles and squares represent the results of the BD simulation
        for $M_x/M_{\rm sat}$ and $M_y/M_{\rm sat}$, respectively,
        while solid and
        dashed lines are the corresponding result of the EFA.
\item{\bf Fig. 6}
        Deviation of normalized magnetization $M_x/M_{\rm sat}$
        (circles) and $M_y/M_{\rm sat}$ (squares) calculated
        from BD simulation and the EFA for the same conditions as in
        Fig.~\ref{fig_stepHstepshear_BDvsEFA}.
        The solid line is the defect of invariance
        as calculated from the matrix norm of Eq.~(\ref{uudefect}).
        For better visibility,
        the matrix norm was multiplied by a factor $0.1$.
\item{\bf Fig. 7}
        Magnetization dynamics as a function of reduced time
        $t/\tau$ for the same conditions as in
        Fig.~\ref{fig_stepHstepshear_BDvsEFA}.
        Circles and squares represent the result of full BD simulation,
        the dashed-dotted line corresponds to the EFA and full lines are the
        result of the combined integration scheme, where the EFA is
        integrated within the boxed regions while otherwise BD simulations
        are performed.
\item{\bf Fig. 8}
        Magnetization dynamics as a function of reduced time
        $t/\tau$ for inception of oscillatory shear flow with
        frequency $\tau\omega=2$ and
        amplitude $\gamma_0=0.5$.
        The magnetic field is oriented in flow direction with $h=2$.
        Circles and squares represent the results of the BD simulation
        for $M_x/M_{\rm sat}$ and $M_y/M_{\rm sat}$, respectively,
        while solid and
        dashed lines are the corresponding result of the EFA.
\item{\bf Fig. 9}
        Deviation of normalized magnetization $M_x/M_{\rm sat}$ (circles)
        and $M_y/M_{\rm sat}$ (squares) calculated by BD simulation and
        from EFA as functions of time $t/\tau$.
        Also shown is the norm of defect of invariance $\|\Delta\|$
        (solid line), which is multiplied by a factor $0.1$ for better
        visibility.
        The same flow conditions as in Fig.~\ref{fig_oscillshear_BDvsEFA}
        are considered.
\item{\bf Fig. 10}
        Magnetization dynamics as a function of reduced time
        $t/\tau$ for the same conditions as in
        Fig.~\ref{fig_oscillshear_BDvsEFA}.
        Symbols are the result of the BD simulation. Solid and dashed lines
        correspond to the combined integration, where the EFA is
        integrated within the boxed regions (indicated by the shading in the 
	upper part) and the BD simulation
        is preformed outside.
\item{\bf Fig. 11}
        Magnetization dynamics as a function of reduced time
        $t/\tau$ for the same conditions as in
        Fig.~\ref{fig_oscillshear_BDvsEFA}, but where the shear flow was
        stopped at time $t=6.5\tau$.
        Symbols are the result of the BD simulation. Solid and dashed lines
        correspond to the combined integration, where the EFA is
        integrated within the boxed regions (indicated by the shading in the 
	upper part) and the BD simulation
        is preformed outside. Dashed-dotted lines are the result of
        the EFA.
\item{\bf Fig. 12}
        Relative error $R_x$ defined in the text as a function of
        CPU time in seconds on a logarithmic scale.
        The same conditions as in
        Fig.~\ref{fig_oscillshear_BDvsEFA} are considered.
        The number above the filled symbols are the
        corresponding values of $\epsilon$.
        Solid lines are guides to the eye.
        Different values of $\epsilon$ decreasing from
        $\epsilon=1$ (EFA) to  $\epsilon=0$ (BD simulation)
        have been chosen in the combined integration scheme
        in order to obtain increasingly more accurate
        results for $M_x$.
\end{description}
\clearpage
\begin{figure}[h]
        \setlength{\unitlength}{1cm}
        \begin{picture}(10,5)
          \thinlines
          \put(2.9,1){\vector(1,0){9.2}}
          \thinlines
          \put(2.9,4){\vector(1,0){9.2}}
          \put(1.2,3.9){macro}
          \put(1.2,0.9){micro}
          \linethickness{0.5mm}
          \put(2.9,1){\vector(1,0){2.6}}
          \put(5.5,1){\vector(0,1){3.0}}\thicklines
          \linethickness{0.5mm}
          \put(5.5,4){\vector(1,0){3.2}}
          \linethickness{0.5mm}
          \put(8.7,4){\vector(0,-1){3.0}}\thicklines
          \put(8.7,1){\vector(1,0){2.5}}
          \put(3.3,0.3){$\|\Delta\|>\epsilon$}
          \put(6.5,0.3){$\|\Delta\|<\epsilon$}
          \put(9.3,0.3){$\|\Delta\|>\epsilon$}
          \put(2.9,4.2){$t_0$}
          \put(5.5,4.2){$t_1$}
          \put(8.7,4.2){$t_2$}
          \put(12.3,3.9){$t$}
          \put(12.3,0.9){$t$}
        \end{picture}
        \caption{ \label{fig_hybridscheme}
          }
\end{figure}

\clearpage
%
%
\begin{figure}[h]
        \setlength{\unitlength}{1cm}
        \begin{picture}(6,6)
        \put(0,0){\centerline{\includegraphics[width=10cm]{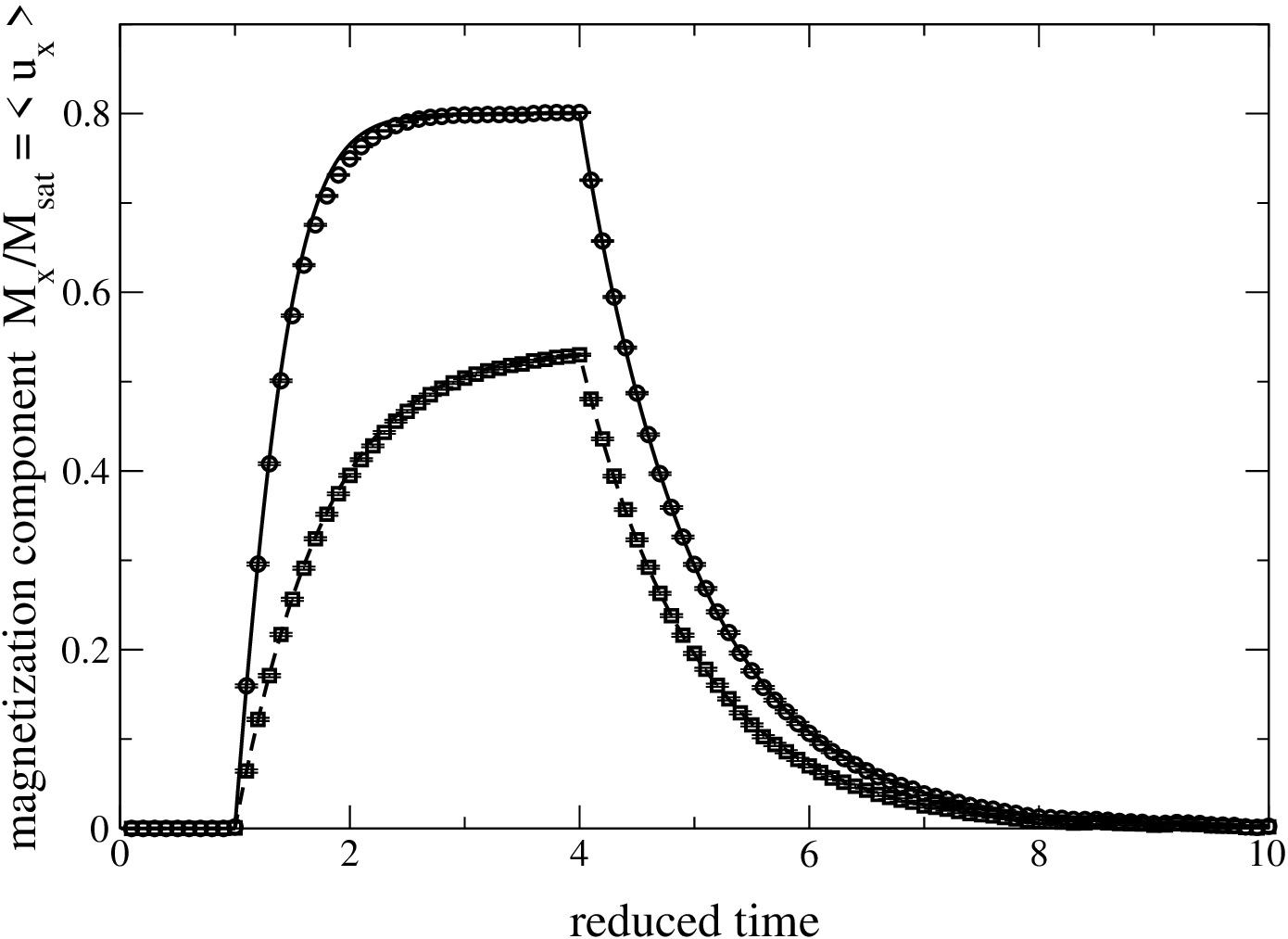}}}
        \put(2.1,3.9){\centerline{\includegraphics[width=3.5cm]{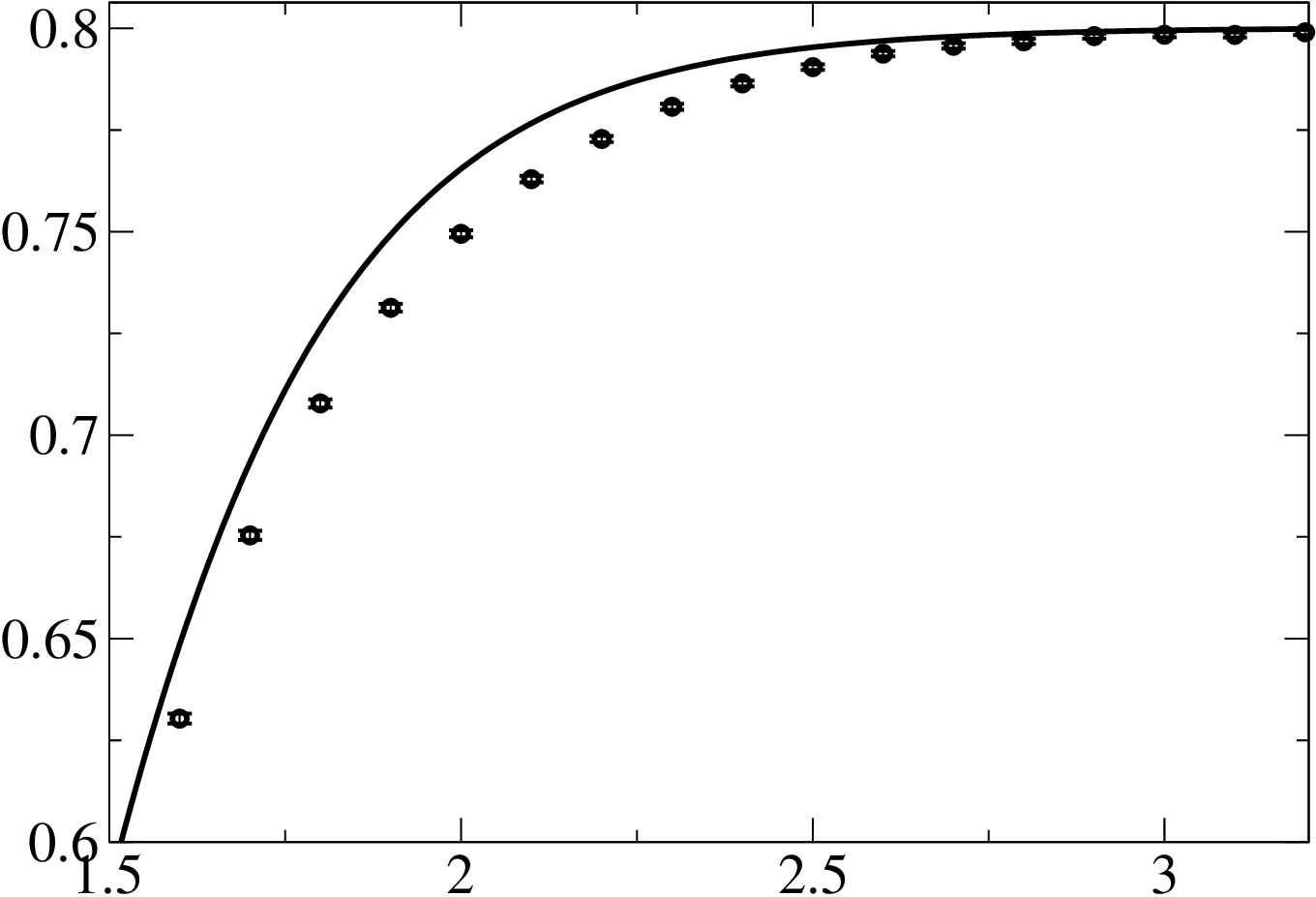}}}
\end{picture}
     \caption[] { \label{fig_stepH_BDvsEFA}
        }
\end{figure}

\clearpage
%
%
\begin{figure}[h]
        \setlength{\unitlength}{1cm}
        \begin{picture}(6,6)
        \put(0,0){\centerline{\includegraphics[width=10cm]{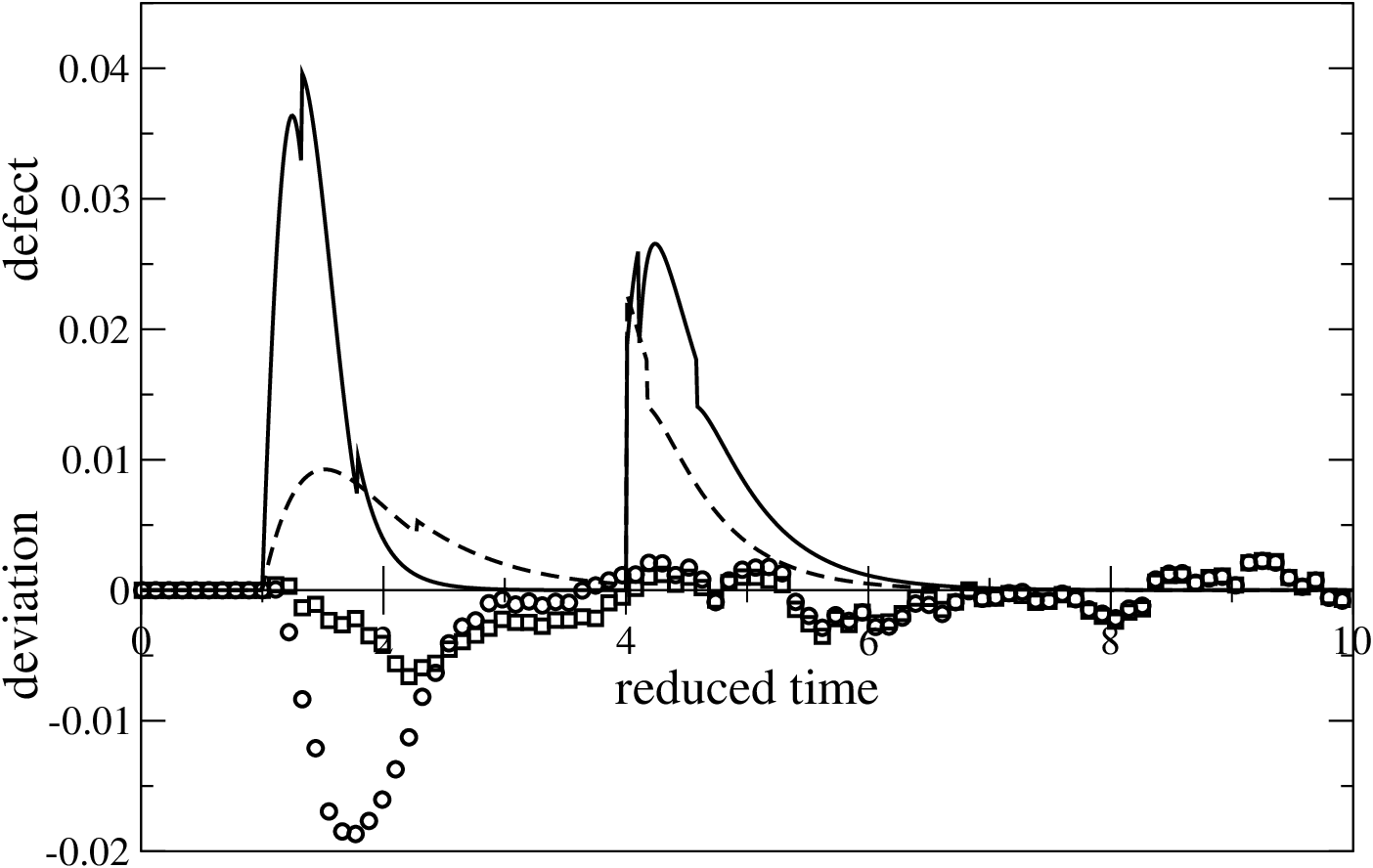}}}
\end{picture}
     \caption[] { \label{fig_diffanddefect_stepH}
        }
\end{figure}

\clearpage
%
%
\begin{figure}[h]
        \setlength{\unitlength}{1cm}
        \begin{picture}(6,6)
        \put(0,0){\centerline{\includegraphics[width=10cm]{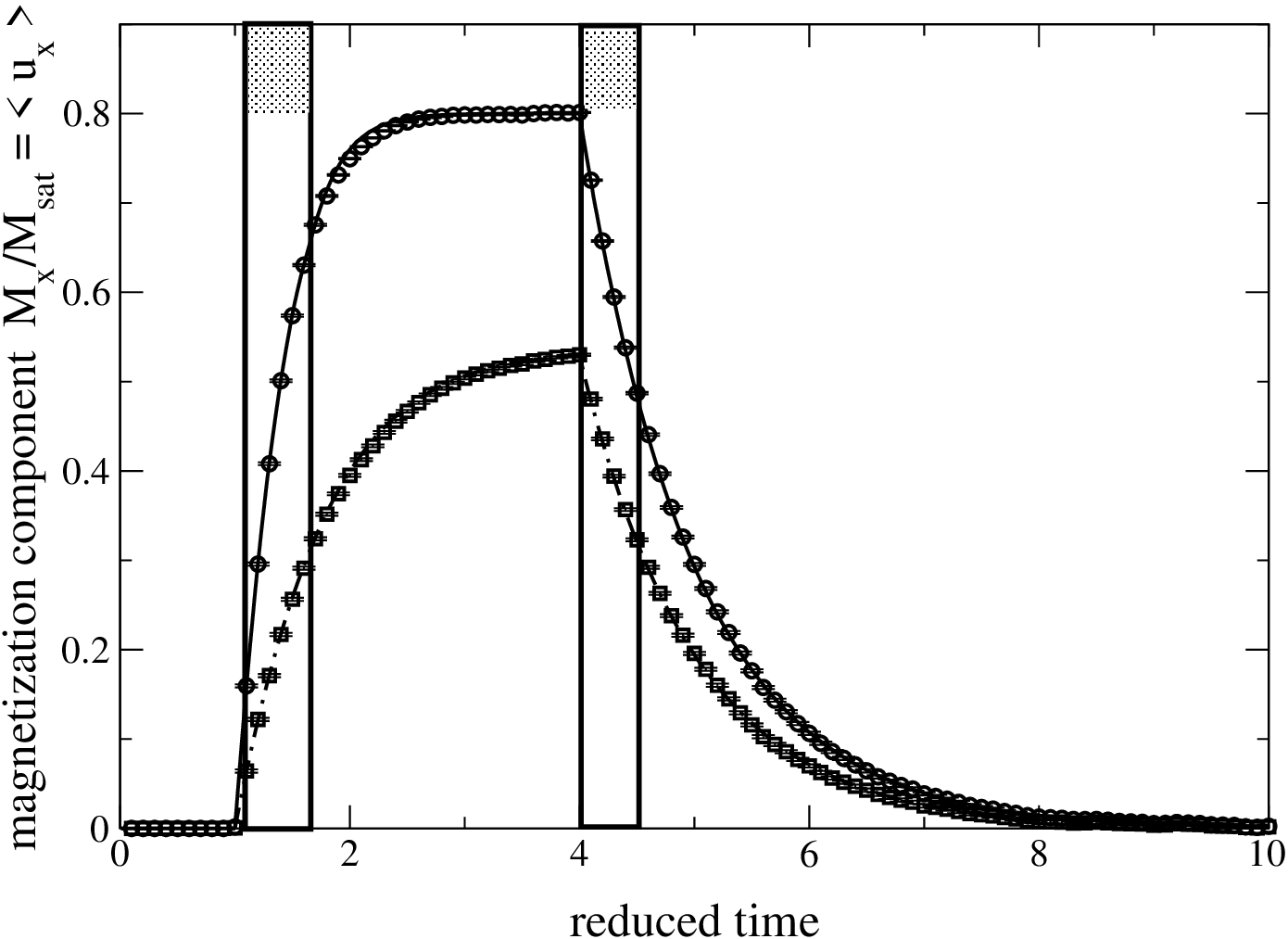}}}
        \put(2.1,3.9){\centerline{\includegraphics[width=3.5cm]{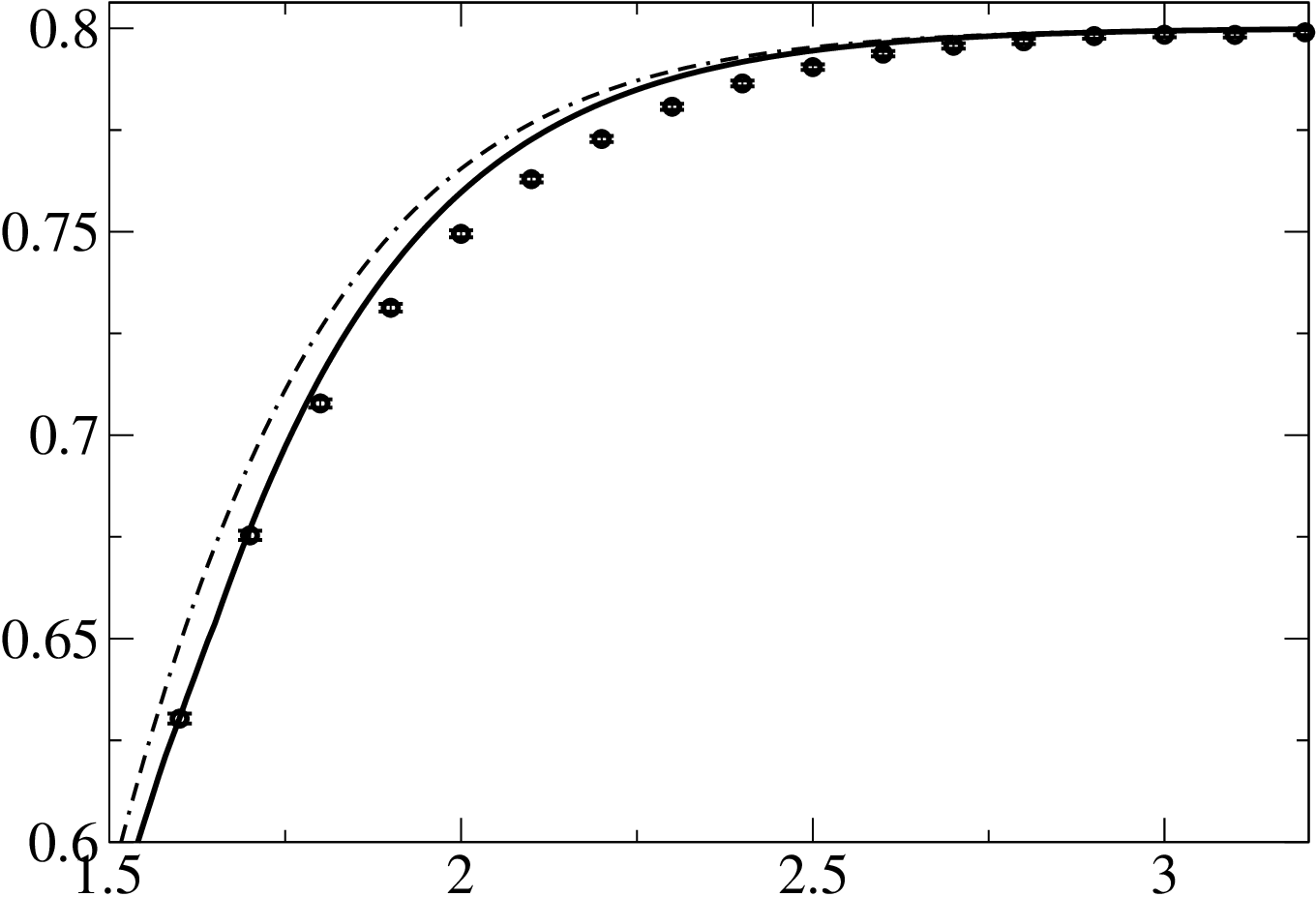}}}
\end{picture}
     \caption[] { \label{fig_stepH_hybridvsBDvsEFA}
        }
\end{figure}

\clearpage
%
%
\begin{figure}[h]
        \setlength{\unitlength}{1cm}
        \begin{picture}(6,6)
        \put(0,0){\centerline{\includegraphics[width=10cm]{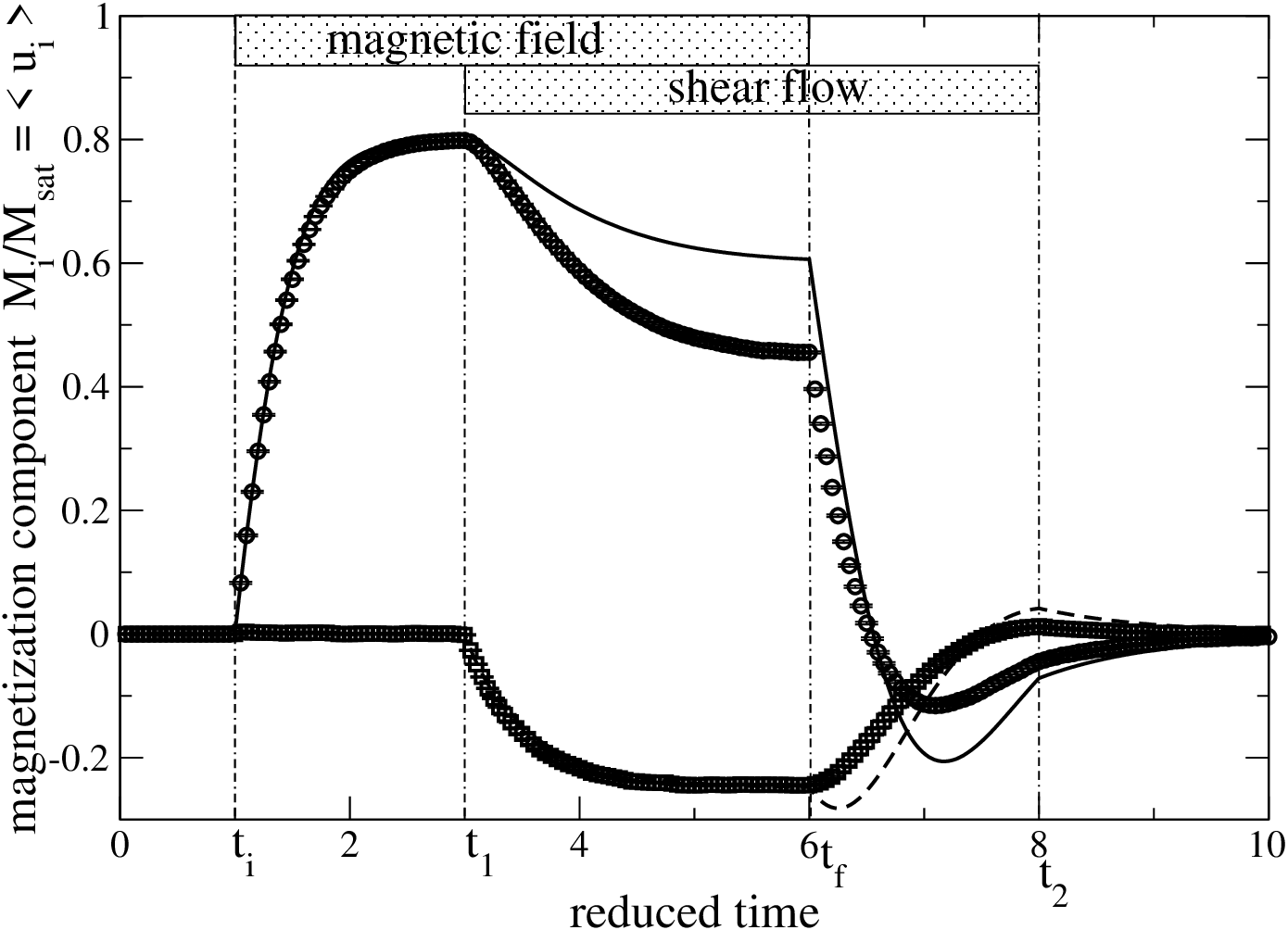}}}
\end{picture}
     \caption[] { \label{fig_stepHstepshear_BDvsEFA}
        }
\end{figure}

\clearpage
%
%
\begin{figure}[h]
        \setlength{\unitlength}{1cm}
        \begin{picture}(6,6)
        \put(0,0){\centerline{\includegraphics[width=10cm]{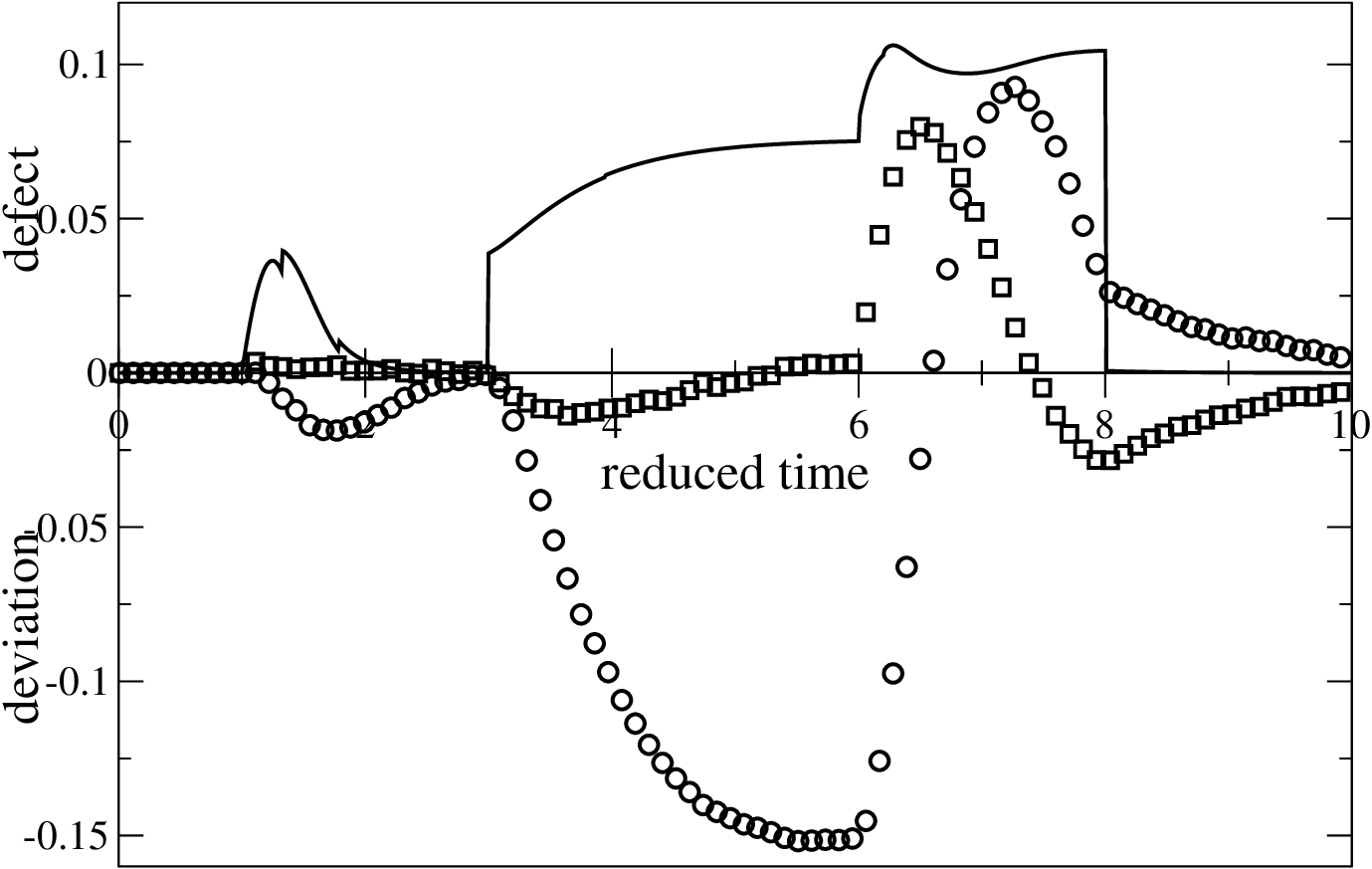}}}
\end{picture}
     \caption[] { \label{fig_diffanddefect_stepHstepshear}
        }
\end{figure}

\clearpage
%
%
\begin{figure}[h]
        \setlength{\unitlength}{1cm}
        \begin{picture}(6,6)
        \put(0,0){\centerline{\includegraphics[width=10cm]{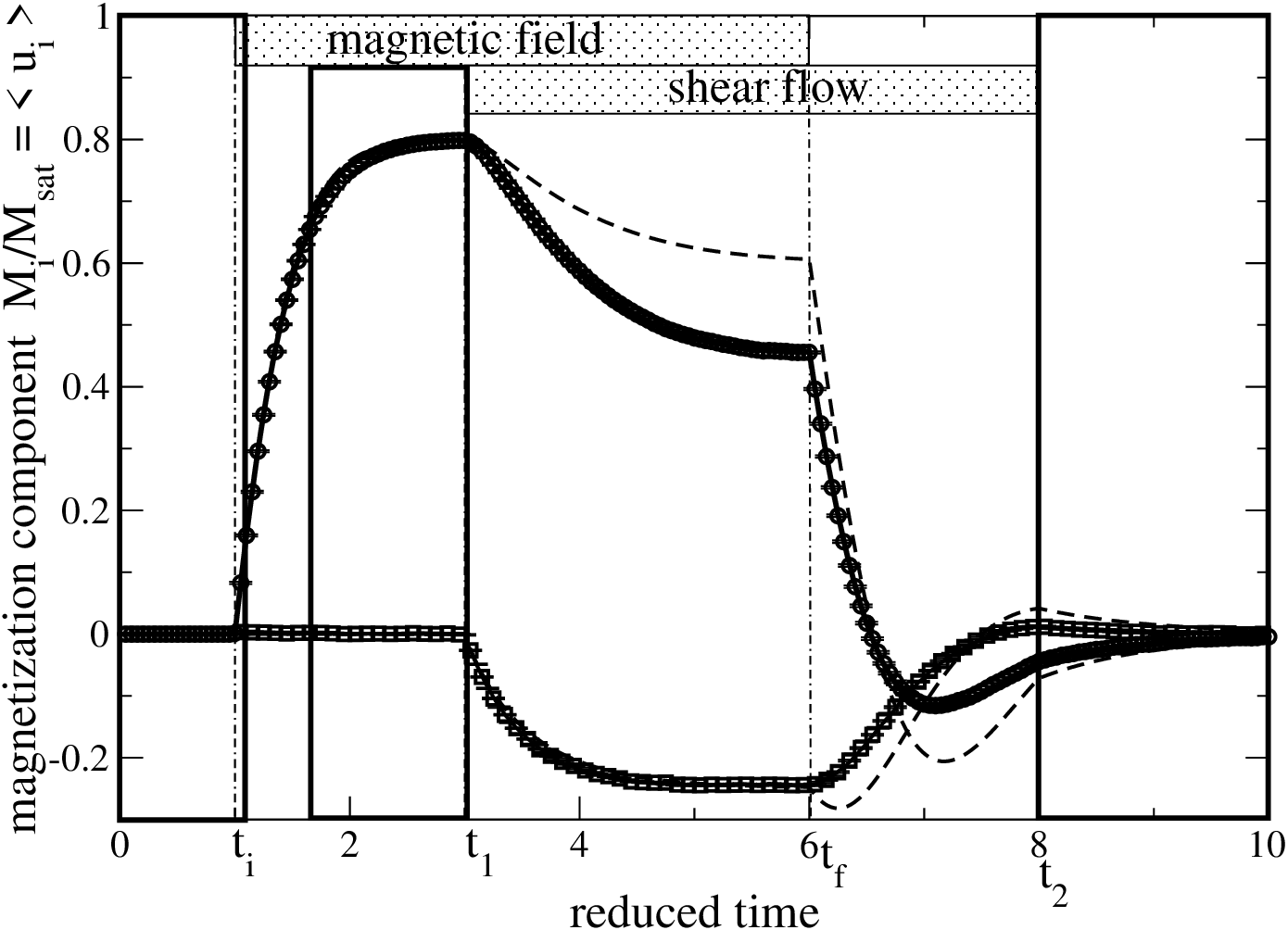}}}
\end{picture}
     \caption[] { \label{fig_stepHstepshear_hybridvsBD}
        }
\end{figure}

\clearpage
%
%
\begin{figure}[h]
        \setlength{\unitlength}{1cm}
        \begin{picture}(6,6)
        \put(0,0){\centerline{\includegraphics[width=10cm]{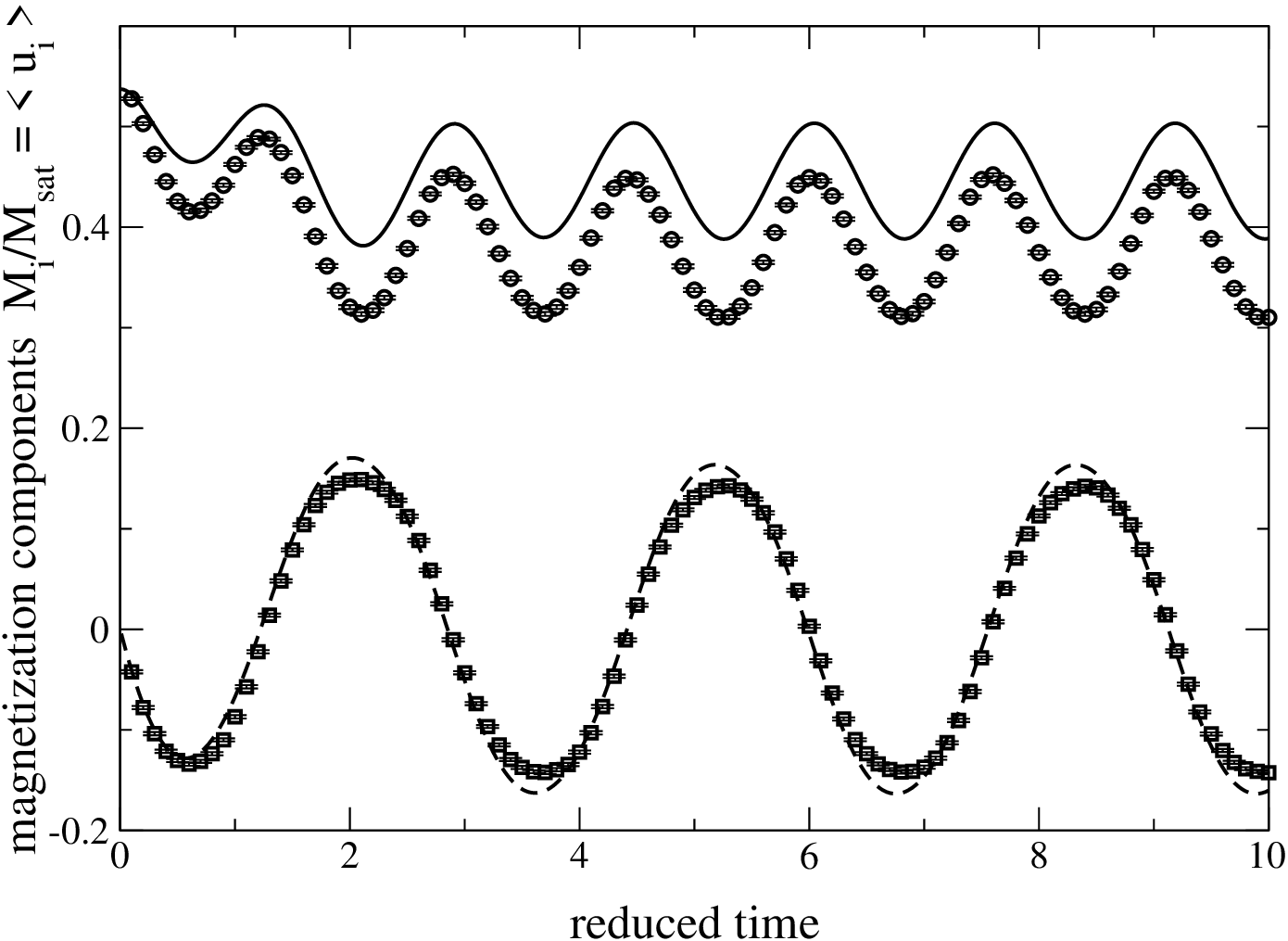}}}
\end{picture}
     \caption[] { \label{fig_oscillshear_BDvsEFA}
        }
\end{figure}

\clearpage
%
%
\begin{figure}[h]
        \setlength{\unitlength}{1cm}
        \begin{picture}(6,6)
        \put(0,0){\centerline{\includegraphics[width=10cm]{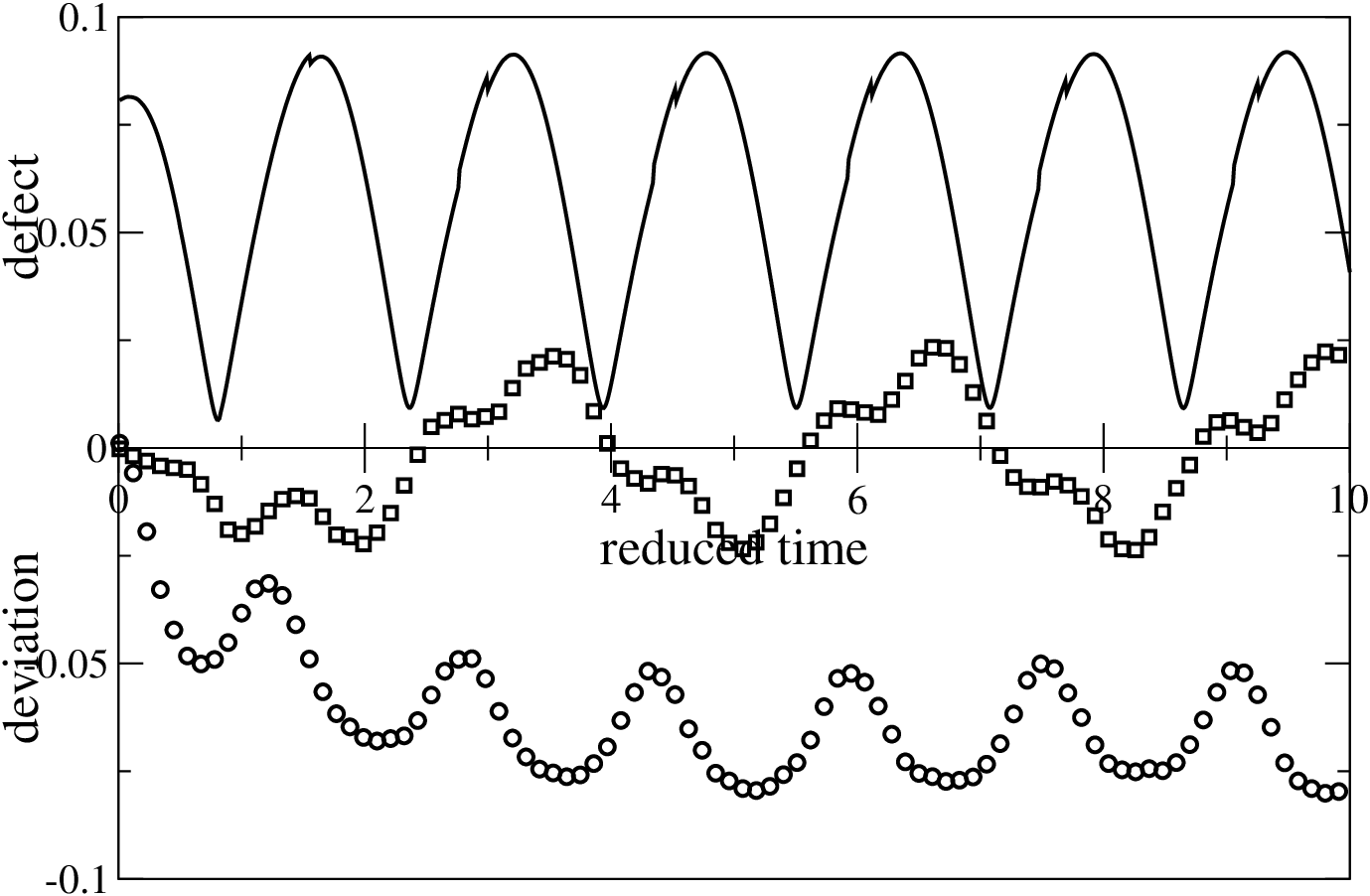}}}
\end{picture}
     \caption[] { \label{fig_diffanddefect_oscillshear}
        }
\end{figure}

\clearpage
%
%
\begin{figure}[h]
        \setlength{\unitlength}{1cm}
        \begin{picture}(6,6)
        \put(0,0){\centerline{\includegraphics[width=10cm]{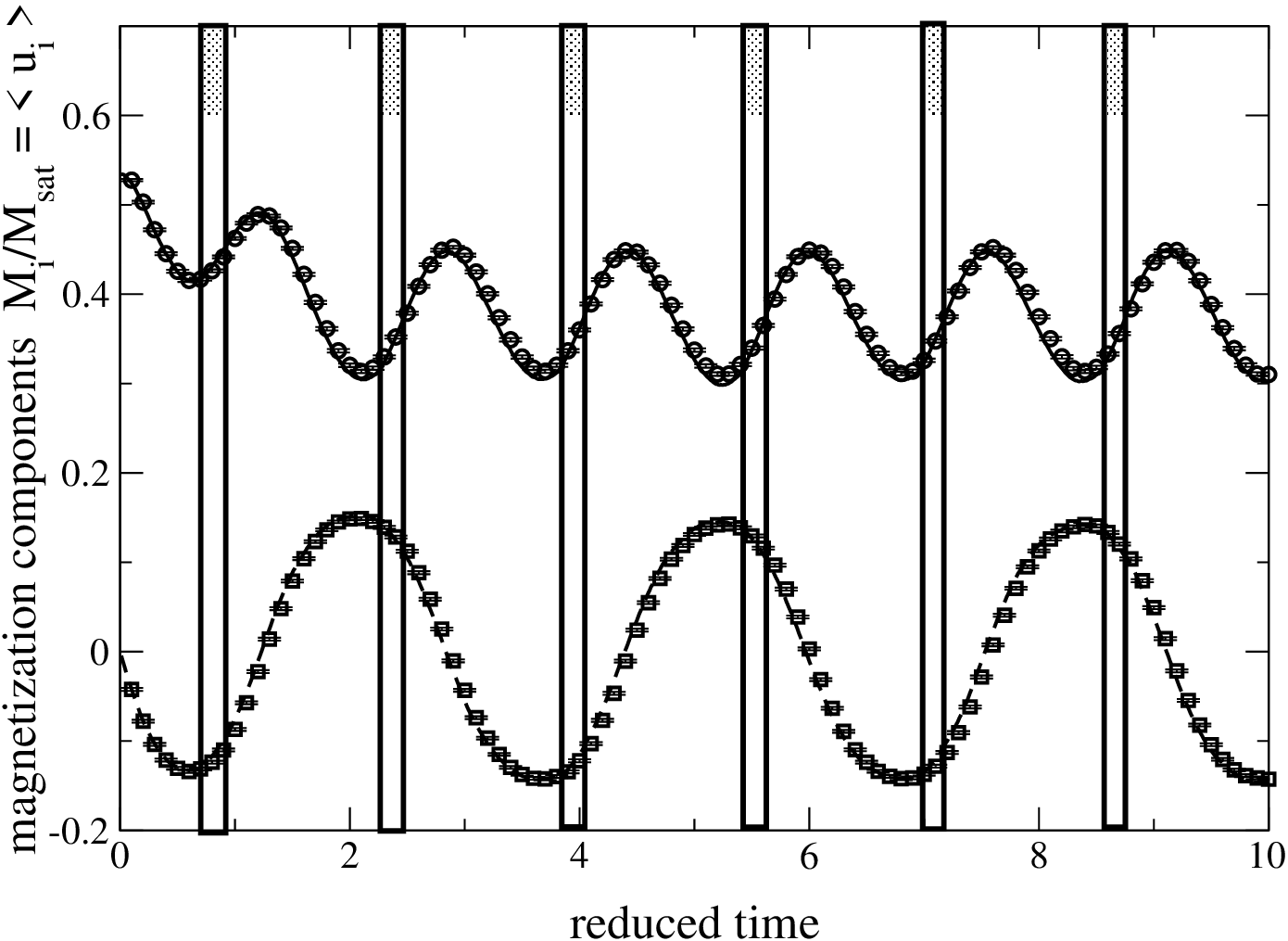}}}
\end{picture}
     \caption[] { \label{fig_oscilshear_hybridvsBDvsEFA}
         }
\end{figure}

\clearpage
%
%
\begin{figure}[h]
        \setlength{\unitlength}{1cm}
        \begin{picture}(6,6)
        \put(0,0){\centerline{\includegraphics[width=10cm]{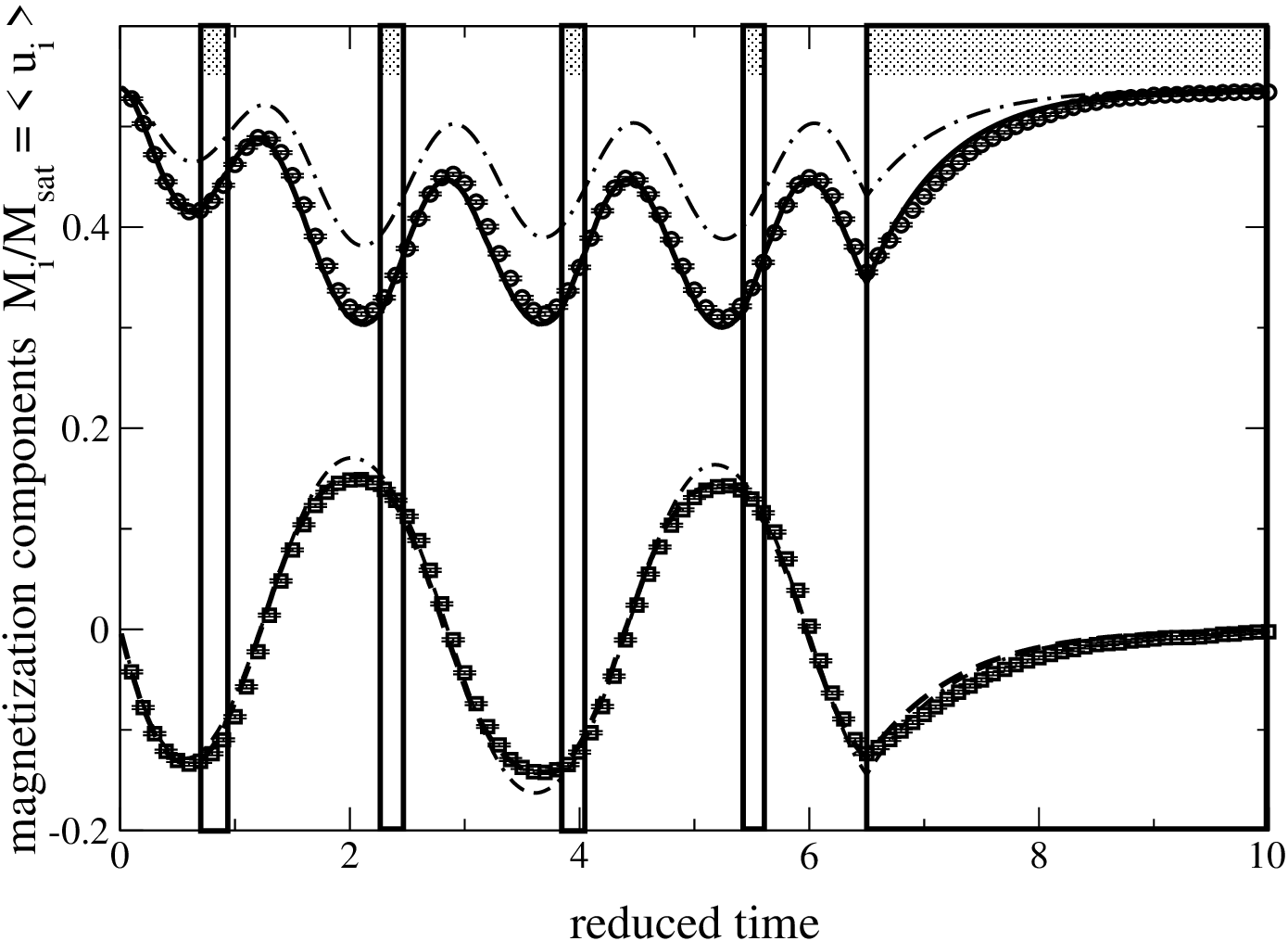}}}
\end{picture}
     \caption[] { \label{fig_oscilshear_Tsteop_hybridvsBDvsEFA}
        }
\end{figure}

\clearpage
%
%
\begin{figure}[h]
        \setlength{\unitlength}{1cm}
        \begin{picture}(6,6)
        \put(0,0){\centerline{\includegraphics[width=10cm]{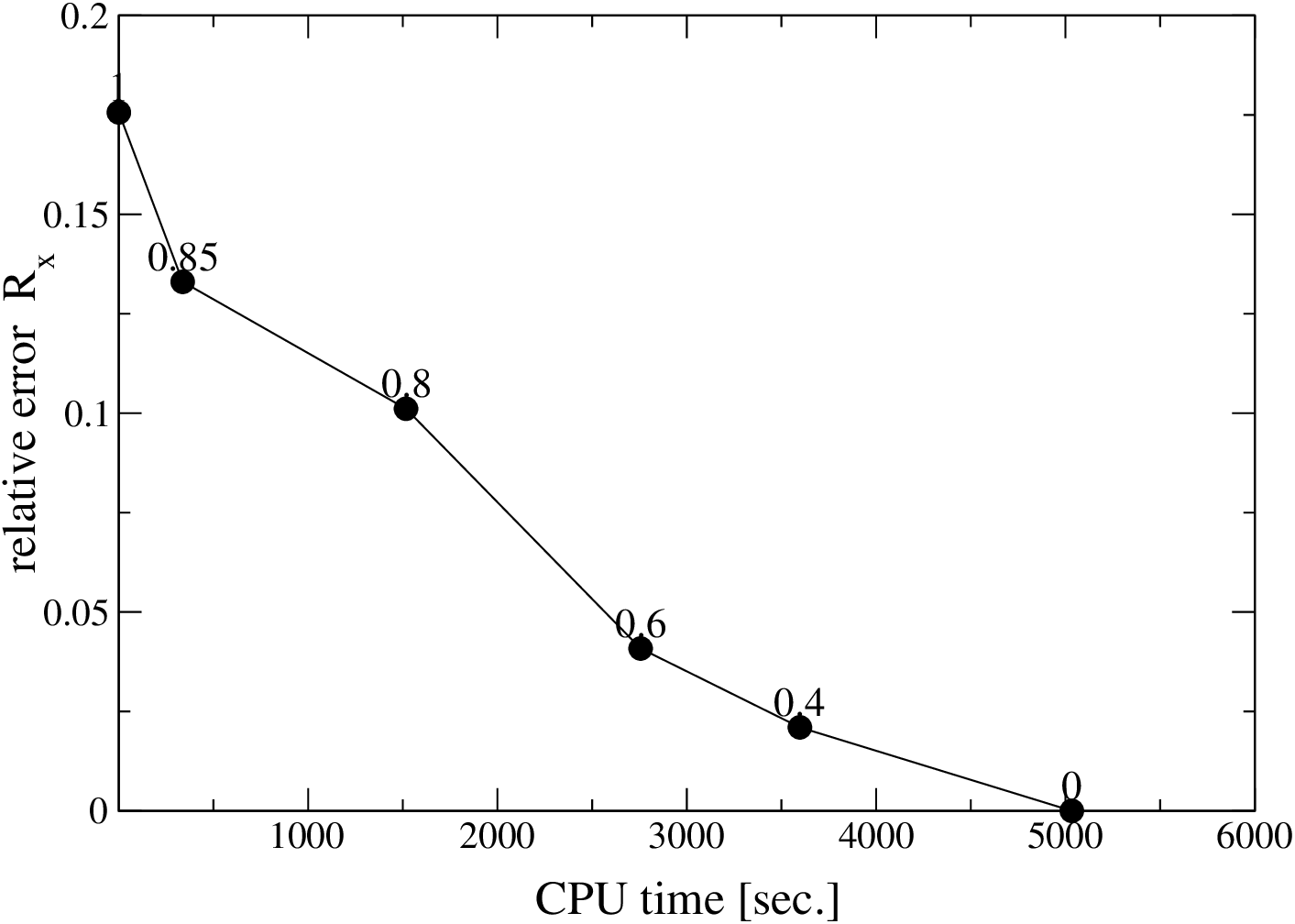}}}
\end{picture}
     \caption[] { \label{fig_error_oscilshear}
        }
\end{figure}

\end{document}